# A Liouville equation for systems which exchange particles with reservoirs: transport through a nano-device.

**Igor V. Ovchinnikov and Daniel Neuhauser**

*Department of Chemistry and Biochemistry, University of California Los Angeles CA 90095-1569 USA.*

**Abstract**. A Redfield-like Liouville equation for an open system that couples to one or more leads and exchanges particles with them is derived. The equation is presented for a general case. A case study of time-dependent transport through a single quantum level for varying electrostatic and chemical potentials in the leads is presented. For the case of varying electrostatic potentials the proposed equation yields, for the model study, the results of an exact solution.

**Introduction**

The problem of transport through mesoscopic or nano-devices has attracted much attention lately [1-15]. Fundamentally, it is quite similar to the problem of an interaction of a system with a bath of "other" particles, whether harmonic or not. One of the simplest and most general approaches for system-bath interactions is the Redfield-Davies approach [16-17], more generally known as the master-equation technique [10,13,18-24], which is suitable for particles which interact with weak baths. The greatest strength of the Redfield-Davies approach is that it allows the solution of complicated systems, with large baths.

Transport, however, is special. Particle transport inherently is associated with open baths (*i.e.*, with the transport of particles from one bath to the other). In this work we therefore consider the equivalent of Redfield's equation for an open system which exchanges particles with reservoirs: electronic transport through a nano-device (for an alternate, partial-trace-free approach, see Ref.[15]). We derive an equation for the one-particle density matrix which has an added source term. We label this equation source-Redfield.

The source-Redfield equation is suitable both for time-independent transport and for time-dependent studies; moreover, it can be extended to include other effects, such as temperature, pressure, or other dissipative mechanisms as it can be combined with the original Redfield-Davies theory in order to take into account interaction with bosonic "heat" baths. A feature which is not used here but would be studied in further work is the extension of the method to deal with the time-evolution of the two-body density matrix, which would allow the studies of more complicated systems (where "system" is used generally to refer to the sub-system, *i.e.*, the device between the leads). Obviously, the formalism is also valid for other cases, such as the interaction of a small adsorbate with a single lead (*e.g.*, with a surface of a crystal or a metal).

In Section II we derive the theory for a general case of leads with varying chemical potentials and/or varying electrostatic potentials (*i.e.*, varying leads' populations and/or varying leads' energies). Section III considers the kinetic limit of diagonal density matrix for the system, and compares, for the case of a system with a single quantum level, the results to that of a previous exact treatment. Discussion follows in Section IV.

**II. Derivation**

The Hamiltonian governing the total system (system+leads) is a combination of non-interacting lead and system Hamiltonians together with a coupling part, $\hat{T}$:

$$\hat{H} = \sum_\alpha \hat{H}_\alpha + \hat{H}_M + \hat{T}, \quad (1)$$

where $\alpha$ is an index over the leads. Typically, only two leads would be used, denoted here as L and R, but the theory is valid for any number of leads, including a single one for an adsorbate-surface study. The central meso- or nano-system can be a single molecule or a more complicated entity.

The Hamiltonian of the system is the sum of a one-body and two-body terms.

$$\hat{H}_M = \hat{H}_{0M} + \hat{U}, \quad (2)$$

where

$$\hat{H}_{0M} = \sum_n E_n \hat{\psi}_n^\dagger \hat{\psi}_n, \tag{3}$$

and $\hat{\psi}_n$ is the electron destruction operator in state *n* of the system; $E_n$ is the "bare" energy of state *n*. There could be various choices for how the two-body interaction $\hat{U}$ is defined which would be explored in future work.

The reservoir Hamiltonian has the form:

$$\sum_\alpha \hat{H}_\alpha = \sum_\alpha \hat{H}_{0\alpha} + \sum_\alpha V_\alpha(t) \hat{N}_\alpha,$$
$$\hat{H}_{0\alpha} = \sum_{A \subset \alpha} E_A \hat{r}_A^\dagger \hat{r}_A, \tag{4}$$

where we introduced the reservoir operators, $\hat{r}_A$, the particle number operator for each lead

$$\hat{N}_\alpha = \sum_{A \subset \alpha} \hat{r}_A^\dagger \hat{r}_A, \tag{5}$$

and the leads' electrostatic potentials $V_\alpha(t)$ (since only the energy difference matters, the electrostatic potential of the system can always be set at zero); A is an index running over all the quantum states in all the reservoirs (*i.e.*, leads); $E_A$ is the energy of state A. The quantum numbers A are not necessarily indices of plane-waves, and they could be, *e.g.*, indices of Bloch states for periodic but non-homogenous leads.

Finally, the coupling between the leads and the Hamiltonian in the Schroedinger picture is:

$$\hat{T} = \sum_{n,A} \left( g_{n,A} \hat{\psi}_n^\dagger \hat{r}_A + h.c. \right), \tag{6}$$

where $g_{n,A}$'s are coupling constants. Henceforth, we will use an interaction picture, with the leads' Hamiltonian plus the one-body part of the system Hamiltonian,

$$\hat{H}_0 = \sum_\alpha \hat{H}_{0\alpha} + \hat{H}_{0M}, \tag{7}$$

as the underlying zeroth order Hamiltonian. In the interaction representation the two-body interaction operator becomes time-dependent $\hat{U} \to \hat{U}(t)$ and the tunneling operator becomes:

$$\hat{T}(t) = \sum_{n,A} \left( g_{n,A} \hat{\psi}_n^\dagger(t) \hat{r}_A(t) + h.c. \right), \tag{8}$$

where

$$\hat{\psi}_n(t) = \hat{\psi}_n e^{-iE_n t},$$
$$\hat{r}_A(t) = \hat{r}_A e^{-iE_A t - i\int_{-\infty}^t V_\alpha(t'') dt''}. \tag{9}$$

The equation governing the (reduced) density matrix of the system, $\hat{\rho}$, is derived similarly to the Redfield-Davies equation, but interaction of the system with bosonic baths is replaced by the term $\hat{T}$ describing the tunneling processes in and out of the reservoirs. The solution of the Liouville equation,

$$i \frac{d \hat{\rho}_T(t)}{dt} = \left[ \hat{T}(t) + \hat{U}(t), \hat{\rho}_T(t) \right], \tag{10}$$

(using $\hbar = 1$) for the full density matrix of the total system, $\hat{\rho}_T$, obeys the relation:

$$\hat{\rho}_T(t) = \hat{\rho}_T^0 - i \int_{-\infty}^t dt' \left[ \hat{T}(t') + \hat{U}(t'), \hat{\rho}_T(t') \right], \tag{11}$$

where $\hat{\rho}_T^0 \equiv \hat{\rho}_T(-\infty)$. The relation can be recast by iterating it as:

$$\hat{\rho}_T(t) = \hat{\rho}_T^0 - i \int_{-\infty}^t dt' \left( \left[ \hat{U}(t'), \hat{\rho}_T(t') \right] + \left[ \hat{T}(t'), \hat{\rho}_T^0 \right] \right.$$
$$- i \int_{-\infty}^{t'} dt'' \left( \left[ \hat{T}(t'), \left[ \hat{U}(t''), \hat{\rho}_T(t'') \right] \right] \right.$$
$$\left. \left. + \left[ \hat{T}(t'), \left[ \hat{T}(t''), \hat{\rho}_T(t'') \right] \right] \right) \right). \tag{12}$$

Taking the time derivative of Eq. (12) and tracing out the reservoirs' degrees of freedom leads to:

$$\frac{d\hat{\rho}(t)}{dt} = -i \left[ \hat{U}(t), \hat{\rho}(t) \right]$$
$$- i \int_{-\infty}^t dt' \sum_{mn} \left( \Sigma_{nm}^>(t,t') \left[ \hat{\psi}_n^\dagger(t), \hat{\psi}_m(t') \hat{\rho}(t') \right] \right.$$
$$+ \Sigma_{nm}^<(t,t') \left[ \hat{\rho}(t') \hat{\psi}_m(t'), \hat{\psi}_n^\dagger(t) \right]$$
$$+ \left[ \hat{\psi}_m(t), \hat{\rho}(t') \hat{\psi}_n^\dagger(t') \right] \bar{\Sigma}_{nm}^>(t',t)$$
$$\left. + \left[ \hat{\psi}_n^\dagger(t') \hat{\rho}(t'), \hat{\psi}_m(t) \right] \bar{\Sigma}_{nm}^<(t',t) \right). \tag{13}$$

Here the tunneling self-energies are the A-sums of single-reservoir-state self-energies, *e.g.*, $\Sigma_{nm}^> = \sum_A (\Sigma_{nm}^{>,A})$.

The latter are defined as:

$$\Sigma_{nm}^{>,A}(t,t') = \Sigma_{nm}^{r,A}(t,t')(1 - n_A(t')),$$
$$\Sigma_{nm}^{<,A}(t,t') = \Sigma_{nm}^{r,A}(t,t') n_A(t'),$$
$$\bar{\Sigma}_{nm}^{>,A}(t,t') = (1 - n_A(t')) \Sigma_{nm}^{a,A}(t',t),$$
$$\bar{\Sigma}_{nm}^{<,A}(t,t') = n_A(t') \Sigma_{nm}^{a,A}(t',t),$$

where

$$\Sigma_{nm}^{r(a),A}(t,t') = g_{n,A} G_A^{r(a)}(t,t') g_{A,m},$$

and $G_A^{r(a)}(t,t')$ is the retarded (advanced) Green function of reservoir level A:

$$G_A^r(t,t') = -i\theta(t-t') e^{-iE_A(t-t') - i\int_{t'}^t dt'' V_\alpha(t'')},$$
$$G_A^a(t',t) = (G_A^r(t,t'))^*.$$

For later use we also define

$$\Sigma_{nm}^{r(a)}(t,t') = \sum_{A} \Sigma_{nm}^{r(a),A}(t,t').$$

Eq.(13) relies on the assumption that the reservoirs are big enough to neglect the feedback action of the small system on them.

Time dependence enters the problem only through the time variation of reservoirs. This time variation can be physically produced, *e.g.*, by coupling of the reservoirs to other, bigger, external systems or bath fields. In the present theoretical model the reservoirs vary in time through variation of phenomenological parameters describing the reservoirs. Only one set of such parameters is usually used in conductance problems - electrostatic potentials in the reservoirs, $V_\alpha(t)$. We already deal with $V_\alpha(t)$ in the formulation. However, there are other possibilities, such as pressure change, charging the reservoirs, or in general a change in the occupation numbers. Even though the reservoirs are not necessarily in chemical (and thermal) equilibrium, for brevity we refer to variation of $n_A(t)$ as being due to a variation of the reservoirs' chemical potentials, $\mu_\alpha(t)$.

In deriving Eq.(13), the two-operator reservoirs' quantum averages appear. The way in which the reservoirs' quantum averages (defined by an *R* subscript) are simplified can be seen in the following example:

$$Tr\left(\hat{r}_A(t)\hat{r}_{A'}^\dagger(t')\right)_R =$$
$$\delta_{AA'} Tr\left(\hat{r}_A(t)\hat{r}_A^\dagger(t')\right)_R =$$
$$\delta_{AA'}\left(iG_A^r(t,t')\right) Tr\left(\hat{r}_A(t')\hat{r}_A^\dagger(t')\right)_R =$$
$$\delta_{AA'}\left(iG_A^r(t,t')\right)\left(1-n_A(t')\right).$$

Here we assume that the reservoir states are uncorrelated; the role of the retarded Green function is to relate an operator at time instant *t* to the operator at time instant *t'*,

$$\hat{r}_A(t) = \left(iG_A^r(t,t')\right)\hat{r}_A(t');$$

the occupation number $n_A(t')$ must be taken at time instant *t'* because the trace in the integral part of Eq.(13) is made at this moment.

The time evolution of the expectation value *A(t)* of any operator $\hat{A}(t)$ acting in the system's Hilbert space is given as:

$$\dot{A}(t) = -i\left\langle\dot{\hat{A}}(t)\right\rangle - i\left\langle\left[\hat{A}(t),\hat{U}(t)\right]\right\rangle$$
$$-i\int_{-\infty}^{t} dt'\sum_{mn}\left(\Sigma_{nm}^>(t,t')\left\langle\left[A(t),\hat{\psi}_n^\dagger(t)\right]\hat{\psi}_m(t')\right\rangle\right.$$
$$+\Sigma_{nm}^<(t,t')\left\langle\hat{\psi}_m(t')\left[\hat{\psi}_n^\dagger(t),A(t)\right]\right\rangle$$
$$+\left\langle\hat{\psi}_n^\dagger(t')\left[A(t),\hat{\psi}_m(t)\right]\right\rangle\bar{\Sigma}_{nm}^>(t',t)$$
$$\left.+\left\langle\left[\hat{\psi}_m(t),A(t)\right]\hat{\psi}_n^\dagger(t')\right\rangle\bar{\Sigma}_{nm}^<(t',t)\right), \quad (14)$$

where the inner traces are taken at time *t'*. In this work we are interested in the one-particle density matrix for the inner system

$$\rho(t) \equiv \rho_{xy}(t) = \left\langle\hat{\psi}_y^\dagger(t)\hat{\psi}_x(t)\right\rangle. \quad (15)$$

(As mentioned, the formalism would be applied in latter work for the evolution of the two-body density matrix). The Liouville-type equation which results is:

$$\dot{\rho} = -i[h\rho - \rho h] + D, \quad (16)$$

where

$$(h\rho)_{xy}(t) = E_x\rho_{xy}(t) + \sum_m \Xi_{xm}^r(t)\rho_{my}(t)$$
$$+\sum_{mn}\int dt'\Sigma_{xm}^r(t,t')\rho_{mn}(t')(-iG_{ny}^a(t',t)),$$
$$(\rho h)_{xy}(t) = \rho_{xy}(t)E_y + \sum_m \rho_{xm}(t)\Xi_{my}^a(t)$$
$$+\sum_{mn}\int dt'(iG_{xm}^r(t,t'))\rho_{mn}(t')\Sigma_{ny}^a(t',t),$$
$$D_{xy}(t) = i\sum_{mA}\int dt'(\Sigma_{xm}^{r,A}(t,t')(-iG_{my}^a(t',t))$$
$$+(iG_{xm}^r(t,t'))\Sigma_{my}^{a,A}(t',t))n_A(t'). \quad (17)$$

Here $\Xi_{xy}^{r(a)}(t)$ is a retarded (advanced) two-body self-energy due to the two-body interaction $\hat{U}$ which will be studied in a future paper; $G_{xm}^{r(a)}(t,t')$ is a dressed, *i.e.*, exact, retarded (advanced) Green function of the mesoscopic system, which obeys the equation

$$\left(i\frac{\partial}{\partial t} - E_x\right)G_{xy}^r(t,t') - \sum_m \Xi_{xm}^r(t)G_{my}^r(t,t')$$
$$-\int dt''\sum_m \Sigma_{xm}^r(t,t'')G_{my}^r(t'',t') = \delta_{xy}\delta(t-t'). \quad (18)$$

Formally, in Eq.(14) one should use Eq.(9) in order to relate the interaction representation operators $\hat{\psi}_x(t)$ and $\hat{\psi}_x^\dagger(t)$ to $\hat{\psi}_x(t')$ and $\hat{\psi}_x^\dagger(t')$ in averages like

$$\langle\hat{\psi}_y^\dagger(t)\hat{\psi}_x(t')\rangle = \langle\hat{\psi}_y^\dagger(t')\hat{\psi}_x(t')\rangle e^{-iE_y(t'-t)}$$
$$= \sum_m \rho_{xm}(t')(-iG_{my}^{a,0}(t',t)),$$
$$\langle\hat{\psi}_y^\dagger(t')\hat{\psi}_x(t)\rangle = e^{-iE_x(t-t')}\langle\hat{\psi}_y^\dagger(t')\hat{\psi}_x(t')\rangle$$
$$= \sum_m (iG_{xm}^{r,0}(t,t'))\rho_{my}(t'). \quad (19)$$

In other words, one should formally use the bare retarded (advanced) Green function

$$G_{xy}^{r,0}(t,t') = -i\theta(t-t')\delta_{xy}e^{-iE_x(t-t')}.$$

However, relations (19) are correct only for "bare" operators in the interaction representation and do not take into account two-body interactions and interactions with reservoirs, *i.e.*, such effects as level mixing and phase decay. The correct answer can be obtained by substituting the bare retarded and advanced Green functions of the system by dressed ones, *i.e.*, by using the exact relations from the Keldysh approach [25]:

$$\langle\hat{\psi}_y^\dagger(t)\hat{\psi}_x(t')\rangle = \sum_m \rho_{xm}(t')(-iG_{my}^a(t',t)),$$
$$\langle\hat{\psi}_y^\dagger(t')\hat{\psi}_x(t)\rangle = \sum_m (iG_{xm}^r(t',t))\rho_{my}(t'). \quad (20)$$

Eqs.(16)-(17) have the form of a Liouville von-Neumann equation for the one-particle density matrix, with a complex Hamiltonian (due to the complexity of the $\Sigma$-terms in the definition of $h\rho$ and $\rho h$ in Eqs.(17)) and an additional driving (or pumping) source term $D$. The complexity of the Hamiltonian results in attenuation of the one-particle density matrix components since particles leave the system to the reservoirs. The driving term, $D$, accounts for the absorption of particles from the reservoirs. The only term in Eq.(16) depending on leads' populations, $n_A$, is the driving term $D$. The driving term vanishes when the leads are unpopulated ($n_A = 0$), *i.e.*, the leads pump the system with particles only when they possess particles themselves.

The current between lead $\alpha$ and the system can be derived as [10,13]:

$$J^\alpha = -e\frac{\partial}{\partial t}Tr(\hat{N}_\alpha\hat{\rho}_T(t)) = ieTr([\hat{N}_\alpha, \hat{T}(t)]\hat{\rho}_T(t)). \quad (21)$$

Taking $\hat{\rho}_T(t)$ from Eq.(11) and assuming that in the infinite past the contacts were not correlated one arrives at the following expression:

$$J^\alpha = -e\sum_{\substack{l,m,n \\ A\subset\alpha}}\int dt'$$
$$\times \text{Im}(\Sigma_{lm}^{r,A}(t,t')(\delta_{mn}n_A(t') - \rho_{mn}(t'))(-iG_{nl}^a(t',t))).$$

### III. Kinetic Limit

### III.a Derivation

The Liouville equation with the source term combined with the equation for the retarded Green function (Eqs. (16)-(17) and (18)) is the main result of the paper. As soon as an approximate form of the dependence of the two-body self-energy $\Xi$ on the one-particle density matrix $\rho$ is chosen, *i.e.*, $\Xi \equiv \Xi(\rho)$, the equations become self-contained and can be propagated numerically. It is interesting, however, to pursue a further approximation for these equations, in order to obtain an analytical expression. For that, we first ignore two-body interactions, and then apply the somewhat drastic kinetic assumption, *i.e.*, assume that the coupling rate (defined later) is much smaller than the characteristic energy difference within the system. This means that the $E_x$'s determine the largest energy scale in the problem and it is convenient to incorporate the phase evolution associated with them into $\rho$ by considering

$$\tilde{\rho}_{xy}(t) = e^{-iE_x t}\rho_{xy}(t)e^{iE_y t},$$

so that:

$$\dot{\tilde{\rho}} = -i[\tilde{h}\tilde{\rho} - \tilde{\rho}\tilde{h}] + \tilde{D}, \quad (22)$$

where

$$(\tilde{h}\tilde{\rho})_{xy}(t) = \sum_{mn}\int dt'\tilde{\Sigma}_{xm}^r(t,t')\tilde{\rho}_{mn}(t')(-i\tilde{G}_{ny}^a(t',t)),$$
$$(\tilde{\rho}\tilde{h})_{xy}(t) = \sum_{mn}\int dt'(i\tilde{G}_{xm}^r(t,t'))\tilde{\rho}_{mn}(t')\tilde{\Sigma}_{ny}^a(t',t),$$
$$\tilde{D}_{xy}(t) = i\sum_{mA}\int dt'(\tilde{\Sigma}_{xm}^{r,A}(t,t')(-i\tilde{G}_{my}^a(t',t))$$
$$+(i\tilde{G}_{xm}^r(t,t'))\tilde{\Sigma}_{my}^{a,A}(t',t))n_A(t'), \quad (23)$$

and

$$i\frac{\partial}{\partial t}\tilde{G}_{xy}^r(t,t')$$
$$-\int dt''\sum_m \tilde{\Sigma}_{xm}^r(t,t'')\tilde{G}_{my}^r(t'',t') = \delta_{xy}\delta(t-t'), \quad (24)$$

with

$$\tilde{\Sigma}_{xy}^{r(a)}(t,t') = e^{iE_x t}\Sigma_{xy}^{r(a)}(t,t')e^{-iE_y t'},$$
$$\tilde{\Sigma}_{xy}^{r(a),A}(t,t') = e^{iE_x t}\Sigma_{xy}^{r(a),A}(t,t')e^{-iE_y t'},$$
$$\tilde{G}_{xy}^{r(a)}(t,t') = e^{iE_x t}G_{xy}^{r(a)}(t,t')e^{-iE_y t'}.$$

The kinetic limit, or weak coupling limit, is related to the small magnitude of the parameter $\kappa = g^2/\Delta E \ll 1$, where $\Delta E$ is a characteristic spacing between the system levels. Diagonal self-energies give corrections of order $g^2$ while off-diagonal self-energies give corrections of order $g^2\kappa$. Therefore, we can neglect off-diagonal terms in the tunneling self-energy which drives the system and simultaneously neglect off-diagonal terms in the one-particle density matrix because there would be no off-diagonal pumping terms. The remaining diagonal terms in the den-sity matrix, *i.e.*, the populations, are defined as $N_x \equiv \tilde{\rho}_{xx}$. Now the system consists of non-interacting levels, each described totally by its population. Substituting the sums over the leads' quantum states by an integration: $\sum_{A \subset \alpha} \to \int \sigma^\alpha(E) dE$, where $\sigma^\alpha(E)$ is the density of states of lead $\alpha$ at energy $E$, we represent the diagonal self-energies as

$$\tilde{\Sigma}^r_{xx}(t,t') = \sum_\alpha \int dE \sigma^\alpha(E) | g_{x,\alpha E} |^2 (-i) e^{-i(E-E_x)(t-t')-i\int_{t'}^{t} V_\alpha(t'')dt''}.$$

It is easy to show that if $\sigma^\alpha(E) | g_{x,\alpha E} |^2$ is a slowly varying function around $E_x$, then the integration over $E$ gives

$$\tilde{\Sigma}^r_{xx}(t,t') \approx (\Delta E_x - i\Gamma_x/2)\delta(t-t'-0^+),$$

where $\Delta E_x$ is the shift of the $x$-level energy and $\Gamma_x$ is the coupling rate to all the reservoirs

$$\Gamma_x = \sum_\alpha \Gamma^\alpha_x, \quad \Gamma^\alpha_x = 2\pi\sigma^\alpha(E_x) | g_{x,\alpha E_x} |^2.$$

Now the retarded Green function of the system is

$$\tilde{G}^r_{xx}(t,t') = -i\theta(t-t')e^{-i(\Delta E_x - i\Gamma_x/2)(t-t')},$$

and Eq. (22) finally takes the form:

$$\dot{N}_x(t) = -\Gamma_x N_x(t)$$
$$+ \int dt' dE \sum_\alpha \Gamma^\alpha_x F_{x,\alpha E}(t,t') n_{\alpha,E}(t'),$$

$$F_{x,\alpha E}(t,t') = \frac{\theta(t-t')}{\pi} \text{Re}\left(e^{-i(E-\tilde{E}_x - i\Gamma_x/2)(t-t')-i\int_{t'}^{t} V_\alpha(t'')dt''}\right), \quad (25)$$

where $\tilde{E}_x = E_x + \Delta E_x$ are the levels' energies renormalized with respect to the interaction with the reservoirs.

The current between lead $\alpha$ and level $x$ is given as:

$$J^\alpha_x(t) = -e\Gamma^\alpha_x \left( N_x(t) - \int_{-\infty}^{t} dt' dE F_{x,\alpha E}(t,t') n_{\alpha,E}(t') \right). \quad (26)$$

The solution of (25) is:

$$N_x(t) = \int_{-\infty}^{t} dt' e^{-\Gamma_x(t-t')} \int_{-\infty}^{t'} dt'' dE \sum_\alpha \Gamma^\alpha_x F_{x,\alpha E}(t',t'') n_{\alpha,E}(t'').$$
(27)

It is easy to show that in the case of a time-independent reservoir populations, $n_{\alpha,E}(t) \equiv n_{\alpha,E}$, formulae (27) and (26) are exactly the result by Wingreen, Jauho and Meir (Eqs.(10) and (11) of Ref.[11]) obtained within the Keldysh diagrammatic approach.

In the stationary case, the electrostatic and chemical potentials are time-independent and one obtains the usual result for a resonant-tunneling current:

$$J = \sum_x \int dE \frac{e}{2\pi} (n^L_E - n^R_E) \frac{\Gamma^L_x \Gamma^R_x}{(E-E_x)^2 + (\Gamma_x/2)^2}.$$

At equilibrium when $n^L_E = n^R_E = \theta(\mu - E)$ the populations of the levels are:

$$N_x = \int_{-\infty}^{\mu} dE \frac{1}{2\pi} \frac{\Gamma_x}{(E-E_x)^2 + (\Gamma_x/2)^2}.$$

If the chemical potential in the leads is sufficiently "separated" from the levels, *i.e.*, $| E_x - \mu | \gg \Gamma_x, \forall x$, the level populations are $N_x = \theta(\mu - E_x)$, *i.e.*, only levels under the chemical potential are populated, as should be.

**III.b Two Different Ways to Picture Time-Dependent Reservoirs**

Our approach allows the reservoir populations to vary and the next goal it to compare the two possible ways to view the reservoirs – as having varying electrostatic potentials or varying chemical potentials. But first consider the limit of slowly varying chemical or/and electrostatic potentials (slowly with respect to the rate $\Gamma$). The function $F_{x,\alpha E}(t,t')$, defined in Eq.(25), is non-negligible only if $(t-t')\Gamma_x \lesssim 1$. If the electrostatic potential varies slowly on the $1/\Gamma_x$ scale then $F_{x,\alpha E}(t,t')$ can approximately be rewritten as:

$$F_{x,\alpha E}(t,t') = \frac{\theta(t-t')}{\pi} \text{Re}\left(e^{-i(E+V_\alpha(t')-\tilde{E}_x - \Gamma_x/2)(t-t')}\right).$$

Then, assuming that the occupation numbers of the reservoir levels, $n_{\alpha,E}(t'')$, are functions of $E - \mu_\alpha$, $n_{\alpha,E}(t'') \equiv f_\alpha(E - \mu_\alpha, t'')$ (*e.g.*, the case of local thermal and chemical equilibrium), Eq.(25) takes the form:

$$\dot{N}_x(t) = -\Gamma_x N_x(t)$$
$$+ \int dt' dE \sum_\alpha \Gamma^\alpha_x F^{St}_{x,\alpha E}(t,t') n_{\alpha,E-(\mu_\alpha + V_\alpha(t'))}(t'), (28)$$

where $F^{St}_{x,\alpha E}(t,t') = F_{x,\alpha E}(t,t')\big|_{V_\alpha(t)=0}$, *i.e.*, evaluated for the case of stationary reservoir level energies:

$$F^{St}_{x,\alpha E}(t,t') = \frac{\theta(t-t')}{\pi} \text{Re}\left(e^{-i(E-\tilde{E}_x - \Gamma_x/2)(t-t')}\right). \quad (29)$$

We see that (28) corresponds to the case of stationary reservoir levels but varying chemical potentials $\mu_\alpha(t) = \mu_\alpha + V_\alpha(t)$. Therefore, the two approaches differ only in the case of rather rapidly varying reservoirs.

**III.c Non-Stationary Conductance of A Single-Level System**

To compare the two ways of picturing reservoirs we apply the source-Redfield equation to a single level system. We take the same parameters of the system as those considered by Wingreen, Jauho and Meir [17]. The two cases considered are the response of the single-level system to a rectangular pulse and an ac bias. The coupling rates to the L and R reservoirs are the same, $\Gamma^L = \Gamma^R = \Gamma/2$. Both the case of varying chemical potentials and varying electrostatic potentials will be presented.

For a varying chemical potentials the level energies in the leads are constant in time. The energy of the only system's level is set at zero. The formulas for the level population and the currents between the system and the leads take the form:

$$\dot{N}(t) = -\Gamma N(t) + \frac{\Gamma}{2}\int dt' dE F_E(t-t')\sum_\alpha n_{\alpha,E}(t'), \quad (30)$$

$$J^\alpha(t) = -J_0\left(N(t) - \int_{-\infty}^{t} dt'' dE F_E(t-t') n_{\alpha,E}(t')\right), \quad (31)$$

where the constant $J_0 = e\Gamma/2$ and $F_E(t-t')$ is given in Eq.(29) with $\Gamma_x = \Gamma, \tilde{E}_x = 0$. We assume that the chemical potentials are antisymmetric with respect to level position, i.e., $\mu^L(t) = -\mu^R(t)$. The populations in the leads are given as:

$$n_E^{L,R}(t) = \theta(-E) + \delta n_E^{L,R}(t),$$
$$\delta n_E^{L,R}(t) = (\theta(\mu^{L,R}(t) - E) - \theta(-E)).$$

Since $(\delta n_E^L(t') + \delta n_E^R(t'))$ and $F_E(t-t')$ are respectively antisymmetric and symmetric functions of $E$, the integral in the *r.h.s.* of Eq.(30) is time independent, so that the level population does not vary in time. It is also easy to show that the average population of the level equals *1/2*. Consequently, the currents through both boundaries of the single-level system are always the same, $J^L(t) = -J^R(t) = J(t)$. The current can be rewritten as:

$$J(t) = J_0 \int \frac{d\omega}{2\pi} e^{-i\omega t} F_E(\omega) \delta n_E^L(\omega),$$
$$F_E(\omega) = \frac{i}{2\pi}\left(\frac{1}{\omega + E + i\Gamma/2} + \frac{1}{\omega - E + i\Gamma/2}\right).$$

For a rectangular pulse, the left chemical potential is $\mu_{rp}^L(t) = \Delta(\theta(t) - \theta(t-\tau))$, where $\Delta$ and $\tau$ are the pulse magnitude and length respectively. The occupation numbers in the left lead are

$$\delta n_E^L(t) = i\int \frac{d\omega}{2\pi} e^{-i\omega t} \frac{1 - e^{i\omega\tau}}{\omega + i0^+}(\theta(E) - \theta(E - \Delta)),$$

and the current is

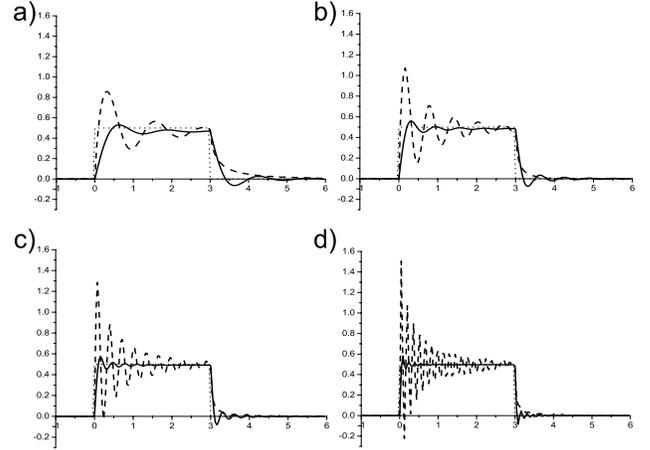

Figure 1: The current $J_{rp}(t)$ through the model single-level system (see text) as a function of time when a rectangular pulse of duration $\tau = 3\Gamma^{-1}$ is applied. Time and current are given in units of $\Gamma^{-1}$ and $J_0 = e\Gamma/2$, respectively. The current is given for four different amplitudes of the pulse $\Delta$: a) $5\Gamma$, b) $10\Gamma$, c) $20\Gamma$, d) $40\Gamma$. Solid and dashed lines represent the current obtained for varying chemical and electrostatic potentials, respectively, and the dotted line is the bias in arbitrary units.

$$J_{rp}(t) = J_0(\theta(t)\chi(t) - \theta(t-\tau)\chi(t-\tau)),$$
$$\chi(t) = \frac{\Gamma}{2\pi}\int_0^\Delta dE \frac{1 - e^{-\Gamma/2 t}(\cos(Et) - 2(E/\Delta)\sin(Et))}{E^2 + (\Gamma/2)^2}.$$

In an ac bias case, the left chemical potential is $\mu_{ac}^L(t) = \Delta(1 - \cos(\omega_0 t))$, where $\Delta$ and $\omega_0$ are the magnitude and frequency of the bias respectively. The populations in the left lead are given as ($2\Delta > E > 0$):

$$\delta n_E^L(t) = \sum_{k=-\infty}^\infty \frac{(-1)^k \sin(a_E k)}{\pi k}\exp(-ik\omega_0 t),$$

where $a_E = \cos^{-1}((E-\Delta)/\Delta)$. The current has the following form:

$$J_{ac}(t) = J_0 \int_0^{2\Delta} dE \sum_{k=-\infty}^\infty 2\frac{(-1)^k \sin(a_E k)}{\pi k} F_E(k\omega_0) e^{-ik\omega_0 t}.$$

For varying electrostatic potentials the currents for a rectangular pulse and an ac bias are the same as obtained in Ref.[11] and are given as:

$$J_{rp}(t) = \frac{J_0}{\pi}\int_{-\infty}^0 dE\,\text{Im}\left(\frac{\Delta e^{(-iE-\Gamma/2)(t-\tau)\theta(t-\tau)}\left(1 - e^{(-i(E+\Delta)-\Gamma/2)\min(t,\tau)}\right)}{(E+\Delta-i\Gamma/2)(E-i\Gamma/2)}\right),$$

$$J_{ac}(t) = -\frac{J_0}{\pi}\text{Im}\left(e^{i\frac{\Delta}{\omega_0}\sin(\omega_0 t)}\sum_{k=-\infty}^\infty J_k\left(\frac{\Delta}{\omega_0}\right)e^{-ik\omega_0 t}\ln\left(1 + i\frac{\Delta - k\omega_0}{\Gamma/2}\right)\right),$$

where $J_k$ is a first-type $k^{th}$-order Bessel function.

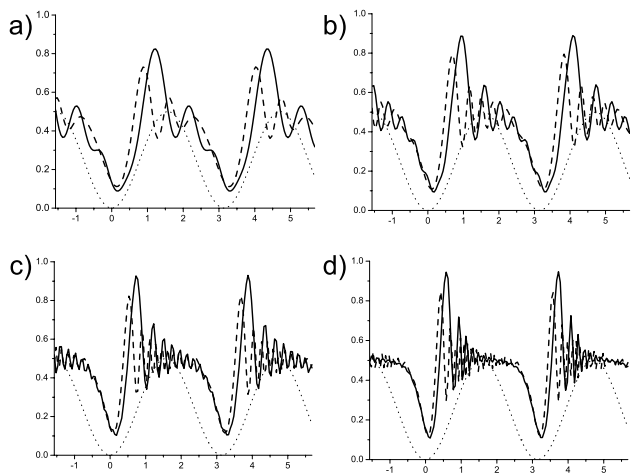

Figure 2: Similar to Fig.1 for the case of an ac bias with frequency $\omega_0 = 2\Gamma$.

The currents $J_{rp}, J_{ac}$ for both rectangular pulse and ac bias are given in Figs.(1) and (2), respectively. The currents are given for four different amplitudes $\Delta = 5\Gamma, 10\Gamma, 20\Gamma$, and $40\Gamma$, and for ac bias $\omega_0 = 2\Gamma$.

Interestingly, the difference between the two currents (due to varying chemical potentials and due to varying electrostatic potentials in the leads) is much more pronounced for a rectangular pulse, while the ac response is similar for both cases. Apparently, this is due to high frequancey components associated with instant switching on and off in the rectangular pulse.

### IV. Discussion and Conclusions

In conclusion, we present a Redfield approach with a source term which is suitable to numerically propagate transport problem under different bias situation, such as time-dependent electrostatic bias, time-dependent charging, or in general time-dependent level energies and level populations in the leads.

The resulting equations are simple to propagate even for complicated systems. They involve a two-time kernel (*i.e.*, $d\rho(t)/dt$ depends on $\rho(t')$ at earlier times), but this can be tracked either by using the slowly varying $\rho(t)$ assumption (*i.e.*, first Markov approximation, $\rho(t) \approx \rho(t')$), or by using more sophisticated approaches (see, *e.g.*, Ref.[24]). Studies using this equation will be presented in future publications.

### Acknowledgements

We are grateful for discussions with Profs. Roi Baer and Ronnie Kossloff. This work was supported by the NSF and the PRF.